\documentclass[twocolumn,twoside]{IEEEtran}

\usepackage{amsmath,amssymb}
\usepackage{epsf}

\newcommand{\C}{{\mathbb C}}
\newcommand{\F}{{\mathbb F}}
\newcommand{\Z}{{\mathbb Z}}
\newcommand{\Hilb}{{\mathcal H}}

\def\ket#1{|#1\rangle}
\def\bra#1{\langle#1|}
\def\braket#1#2{\langle#1|#2\rangle}
\def\bm#1{#1}
\def\transp{{\texttt t}}
\def\w{\omega}
\let\ds\displaystyle

\def\BibTeX{{\rm B\kern-.05em{\sc i\kern-.025em b}\kern-.08em
    T\kern-.1667em\lower.7ex\hbox{E}\kern-.125emX}}

\newtheorem{theorem}{Theorem}
\newtheorem{lemma}{Lemma}

\begin{document}

\title{Graphs, Quadratic Forms, and Quantum Codes}

\author{Markus Grassl\thanks{This work was completed while both
M. Grassl and M. R\"otteler were with Institut f\"ur Algorithmen und
Kognitive Systeme (IAKS), Fakult\"at f\"ur Informatik, Universit\"at
Karls\-ruhe (TH), Am Fasanengarten 5, 76\,128 Karlsruhe, Germany
(e-mail: $\{$grassl,roettele$\}$@ira.uka.de).\protect\newline\indent
A. Klappenecker is with the Department of Computer Science, Texas A\&M
University, College Station, TX 77843 USA (e-mail:
klappi@cs.tamu.edu).}, Andreas Klappenecker, and Martin R\"otteler\\[1ex]
\parbox{\hsize}{\centering \normalfont\small 
Paper presented at ISIT 2002, June 30 -- July 5, 2000, Lausanne,
Switzerland (submitted October 19, 2001)}}

\markboth{In: Proceedings of the 2002 IEEE International Symposium on
  Information Theory}{M. Grassl, M. R{\"o}tteler, and A.
  Klappenecker: Graphs, Quadratic Forms, and Quantum Codes}

\maketitle

\begin{abstract}
We show that any stabilizer code over a finite field is equivalent to
a graphical quantum code. Furthermore we prove that a graphical
quantum code over a finite field is a stabilizer code. The technique
used in the proof establishes a new connection between quantum codes
and quadratic forms. We provide some simple examples to illustrate our
results.
\end{abstract}


\section{Graphical Quantum Codes}\label{sec:graphcodes}

Let $A$ be the additive group of a finite field $\F_{p^m}$.  Denote by
$\Hilb$ the complex vector space $\C^{\alpha}$ of dimension
$\alpha=|A|$.  Let $B$ be an orthonormal basis of $\Hilb^{\otimes n}$
consisting of basis vectors $\ket{y}$ labeled by elements of the group
$A^n$.  Let $K\cong A^k$ and $N\cong A^n$ be subgroups of $A^{k+n}$
such that $A^{k+n}=K\times N$.

Following the definition of Schlingemann and Werner in
\cite{schlingemann00}, a \textit{graphical quantum code} is an
$\alpha^k$-dimensional subspace~$Q$ of $\Hilb^{\otimes n}$, which is
spanned by the vectors
\begin{equation}\label{eq:gcode}
\ket{x} = \frac{1}{\sqrt{\alpha^n}}\sum_{y\in N} 
\Bigl(\prod_{\begin{smallmatrix}i,j=1\\i<j\end{smallmatrix}}^{k+n}\chi(z_i,z_j)^{\Gamma_{ij}}\Bigr)
\ket{y},
\end{equation}
where $x\in K$ and $z=x+y\in K\times N\cong A^{k+n}$. The coefficients
on the right hand side are given by the values of a non-degenerate
symmetric bicharacter $\chi$ on $A\times A$. The exponents
$\Gamma_{ij}$ are given by the adjacency matrix $\Gamma$ of a weighted
undirected graph with integral weights, $\Gamma_{ij}\in \Z$. As
(\ref{eq:gcode}) is independent of the diagonal elements
$\Gamma_{ii}$, we can assume without loss of generality  that the graph
has no loops.

In \cite{schlingemann00} the authors raised the question whether or
not every stabilizer code is equivalent to a graphical quantum
code. Our main result gives an affirmative answer to this question:
\begin{theorem}
Any stabilizer code over the alphabet $A=\F_{p^m}$ is equivalent to a
graphical quantum code. Conversely, any graphical quantum code over
$A$ is a stabilizer code.
\end{theorem}

In the sequel, we will prove this theorem. First, we show that any
graphical code over an extension field $\F_{p^m}$ can be reformulated
as a graphical code over the prime field $\F_p$. Then we compute the
stabilizer associated with a graphical code, followed by the
construction of a graphical representation of a stabilizer code. We
conclude by giving examples which illustrate both directions of our
main theorem.

\begin{lemma}
Any symmetric bicharacter $\chi$ over the abelian group $A\cong\F_p^m$
can be written as
\begin{equation}\label{eq:bichar}
\chi(h,g)=\exp\left(\frac{2\pi i}{p} b(h,g)\right),
\end{equation}
where $b$ is a symmetric bilinear form over $\F_p$, i.e., 
$$
b(h,g)=h^\transp M g
$$
where $M$ is a symmetric matrix over
$\F_p$.
\end{lemma}
\begin{proof}
For fixed $h\in A$, the mapping $g\mapsto \chi(h,g)$ is a character of
$A$. Any character $\zeta$ of $A$ can be written as
$\zeta(g)=\exp(2\pi i/p \cdot h^\transp g)$ where $h^\transp g$
denotes the inner product of the group element $g$ identified with a
vector in $\F_p^m$ and the vector $h\in \F_p^m$. As
$\chi(h_1+h_2,g)=\chi(h_1,g)\chi(h_2,g)$ and the group $A$ is
(non-canonically) isomorphic to its character group $A^*$, the
bicharacter $\chi$ can be written as
$$
\chi(h,g)=\exp\left(\frac{2\pi i}{p} (Mh)^\transp g\right),
$$
where $M$ is an $m\times m$ matrix over $\F_p$. Symmetry of the
bicharacter implies symmetry of $M$.
\end{proof}

Using this lemma, eq.~(\ref{eq:gcode}) can be rewritten as
\begin{eqnarray}
\ket{x}& =&\frac{1}{\sqrt{\alpha^n}}\sum_{y\in N} 
\Bigl(\prod_{\begin{smallmatrix}i,j=1\\i<j\end{smallmatrix}}^{k+n}
\exp(2\pi i/p\cdot (z_i^\transp M z_j))^{\Gamma_{ij}}
\Bigr)
\ket{y}\nonumber\\
&=&\frac{1}{\sqrt{\alpha^n}}\sum_{y\in N} 
  \exp\left(\frac{2 \pi i}{p} q(v)\Bigr)\right)\ket{y}.\label{eq:gcode2}
\end{eqnarray}
Here we identify $x+y\in \F_{p^m}^{k+n}$ with $v=(v_i)\in
\F_p^{m(k+n)}$. Furthermore, $q$ is the quadratic form
\begin{equation}\label{eq:quadform}
q(v):=\sum_{\begin{smallmatrix}i,j=1\\i<j\end{smallmatrix}}^{m(k+n)}
\Gamma'_{ij}v_i v_j
\end{equation}
on $\F_p^{m(k+n)}$ defined by the symmetric matrix
$\Gamma':=\Gamma\otimes M$. Hence the states~(\ref{eq:gcode}) of the
graphical quantum code $Q$ can be expressed in the form
\begin{equation}\label{eq:char_quad_code}
\ket{x}=\frac{1}{\sqrt{\alpha^n}}\sum_{y\in N} 
\zeta(q(x+y))\ket{y},
\end{equation}
where $\zeta$ is a non-trivial additive character of $\F_p$ and $q$ is the
quadratic form (\ref{eq:quadform}) on $\F_p^{m(k+n)}$. We will take advantage of this presentation in the following sections.

Finally, identifying the vector space $\C^{p^m}$ with $(\C^p)^{\otimes
m}$ allows us to reformulate~(\ref{eq:gcode}) in the form
\begin{equation}\label{eq:gcode_p}
\ket{x} = \frac{1}{\sqrt{\alpha^n}}\sum_{y\in \F_p^{mn}} 
\Bigl(\prod_{\begin{smallmatrix}i,j=1\\i<j\end{smallmatrix}}^{m(k+n)}
\tilde{\chi}(v_i,v_j)^{\Gamma'_{ij}}
\Bigr)\ket{y},
\end{equation}
where $\tilde{\chi}$ is a (non-trivial) bicharacter on $\F_p$ and
$x\in\F_p^{mk}$, $v=(x,y)\in\F_p^{m(k+n)}$. Therefore, it is
sufficient to study graphical quantum codes over the prime field
$\F_p$.

\section{Orthonormal Basis of a Graphical Quantum Code}
In general, the vectors defined by (\ref{eq:gcode}) need not
form a basis of the graphical quantum code. In this section, we derive
conditions for the bicharacter $\chi$ and the graph $\Gamma$ under
which the vectors form an orthonormal basis of the code. 

From the preceding it is sufficient to consider a graphical quantum
code $Q$ defined over the additive group $A$ of the prime field
$\F_p$. The code is spanned by the vectors $\ket{x}$ according to
(\ref{eq:gcode_p}). We can associate with the quadratic form $q$ of
(\ref{eq:quadform}) a symmetric bilinear form $b$ on $A^{n+k}$ given
by
$$
b(v_1,v_2)=q(v_1+v_2)-q(v_1)-q(v_2).
$$
For $v_1=x\in K$ and $v_2=y\in N$, this implies a block structure of
the adjacency matrix $\Gamma$, namely
\begin{equation}\label{eq:Gamma_block}
\Gamma=\left(
\begin{array}{c|c}
M_x&B\\
\hline
B^\transp & M_y
\end{array}
\right),
\end{equation}
where the symmetric matrices $M_x$ and $M_y$ correspond to the
restriction of the quadratic form $q$ to $K$ and $N$, resp., and the
$k\times n$ matrix $B$ corresponds to the bilinear form
$b(x,y)=x^\transp B y$. From (\ref{eq:char_quad_code}) we get
\begin{eqnarray}
\ket{x}
&=&\frac{1}{\sqrt{\alpha^n}}\sum_{y\in N}\zeta(q(x+y))\ket{y}\nonumber\\
&=&\frac{1}{\sqrt{\alpha^n}}\sum_{y\in N}\zeta(b(x,y)+q(x)+q(y))\ket{y}\nonumber\\
&=&\frac{\zeta(q(x))}{\sqrt{\alpha^n}}\sum_{y\in N}\zeta(b(x,y)+q(y))\ket{y}.\label{eq:code_bin_quad}
\end{eqnarray}
Notice that $\zeta(q(x))$ yields an insignificant phase factor and
thus we can assume without loss of generality that $K$ is totally
isotropic, i.e., $q(x)=0$ for all $x\in K$. For the adjacency matrix
$\Gamma$, this implies $M_x=0$. If $\zeta$ is the trivial character,
the coefficients of the right hand side are independent of $x$. In
this case the code is one-dimensional. Hence we require $\zeta$ to be
non-trivial, say $\zeta(g)=\exp(2\pi i/p\cdot g)$.

The inner product of two base states $\ket{x}$ and $\ket{x'}$ of the code
is given by
\begin{eqnarray*}
\braket{x'}{x}
&=&\frac{1}{|N|}\sum_{y\in N}\overline{\zeta(b(x',y)+q(y))}{\zeta(b(x,y)+q(y))}\\
&=&\frac{1}{|N|}\sum_{y\in N}\zeta(b(x,y)-b(x',y))\\
&=&\frac{1}{|N|}\sum_{y\in N}\zeta(b(x-x',y)).
\end{eqnarray*}
This sum is either zero or one, that is, the vectors are either
orthogonal or identical. The sum vanishes unless
$b(x-x',y)=(x-x')^\transp By=0$ for all $y\in N$, that is, unless
$x-x'$ lies in the kernel of $B$. Imposing orthogonality of different
states, i.e., $\braket{x'}{x}=0$ unless $x'=x$, implies that this
kernel needs to be trivial. In other words, if we view $B$ as a matrix
over $\F_p$ then the rank of this matrix is $k$.

\section{The Stabilizer of a Graphical Quantum Code}\label{sec:stabilizer}
For $\bm{a},\bm{d}\in\F_p^n$, we define the following operators:
\begin{eqnarray*}
X^{\bm{a}}&:=&\sum_{y\in\F_p^n}\ket{y+\bm{a}}\bra{y}\\
\mbox{and}\quad
Z^{\bm{d}}&:=&\sum_{y\in\F_p^n}\w^{d^\transp  y}\ket{y}\bra{y},
\end{eqnarray*}
where $\w\in C$ is a primitive $p^{\text{th}}$ root of unity. The
set of unitary operators ${\cal E} := \{X^{\bm{a}}Z^{\bm{d}} \colon
\bm{a}, \bm{d} \in \F_p^n\}$ is an orthonormal basis for the vector
space of ${p^n\times p^n}$ matrices with respect to the trace inner
product $\langle A|B\rangle = {\rm tr}(A^\dagger B)/p^n$. What is more,
${\cal E}$ defines a {\em nice error basis} \cite{KR:2000}. The group
$G$ generated by ${\cal E}$ is an extra-special $p$-group. The
elements $X^{\bm{a}}Z^{\bm{d}}$ and
$X^{\bm{a^\prime}}Z^{\bm{d^\prime}}$ commute up to scalars as follows
\cite{ashikhmin00}
\begin{equation}\label{eq:commutator}
(X^{\bm{a}}Z^{\bm{d}})(X^{\bm{a'}}Z^{\bm{d'}}) = 
\w^{\langle a,d' \rangle - \langle a', d \rangle}
(X^{\bm{a'}}Z^{\bm{d'}})(X^{\bm{a}}Z^{\bm{d}}),
\end{equation}
where we have used the standard inner product $\langle a, d \rangle :=
\sum_{i=1}^n a_i d_i \in \F_p$.  Equivalently, two elements of the
group $G$ commute if the vectors $(a,d),(a',d')\in\F_p^{2n}$ are
orthogonal with respect to the symplectic inner product \cite{Rains99}
\begin{equation}\label{eq:symplectic}
{\langle a,d' \rangle - \langle a', d \rangle}.
\end{equation}
Hence, the elements of $G$ are given
by $\{\w^\gamma X^{\bm{a}}Z^{\bm{d}}\colon \gamma \in \F_p, \bm{a},
\bm{d} \in \F_p^n\}$.

A {\em stabilizer code} $Q \subseteq \Hilb^{\otimes n}$ with respect
to the nice error basis ${\cal E}$ corresponds to a joint eigenspace
of an abelian normal subgroup of $G$. In order to compute the
stabilizer group of the graphical code, we consider the action of an
operator $\w^\gamma X^{\bm{a}}Z^{\bm{d}}$ on the states
(\ref{eq:char_quad_code}).  Recall that $\w^\gamma X^aZ^d$ belongs to
the stabilizer of the graphical quantum code~$Q$ if and only if
$\w^\gamma X^aZ^d\ket{x}=\ket{x}$ for all $x\in K$. We claim that the
stabilizer of the graphical quantum code $Q$ is given by the following
set of operators
\begin{equation}\label{eq:stabilizer}
S_Q= \{ \w^{q(a)}X^aZ^{a M_y}\,|\, a\in N\;\mbox{such that $Ba=0$}\},
\end{equation}
where $M_y$ denotes the $n\times n$ submatrix of the adjacency matrix
$\Gamma$ defined in (\ref{eq:Gamma_block}). Indeed, using the
character $\zeta(\gamma):=\w^\gamma$ in (\ref{eq:char_quad_code}),
straightforward calculation shows that
\begin{equation}\label{eq:straight} 
\begin{array}[b]{@{}ll@{}}
\w^\gamma X^aZ^d\ket{x}\\
\ds=\frac{1}{\sqrt{\alpha^n}}\sum_{y\in N} 
\zeta(\gamma+b(x,y-a)+ q(y-a)+d^\transp(y-a))\ket{y}.
\end{array}
\end{equation}
Comparing equations (\ref{eq:straight}) and (\ref{eq:code_bin_quad})
yields
$$
\zeta(\gamma+b(x,y-a)+ q(y-a)+d^\transp(y-a))=\zeta(b(x,y)+q(y))
$$
for all $x\in K$ and $y\in N$. As the character $\chi$ is faithful, we
can simplify this to
\begin{equation}\label{eq:inter}
\gamma-b(x,a)-b(a,y)+q(a)+d^\transp(y-a)=0.
\end{equation}
This formula holds for any choice of $x\in K$, thus $b(x,a)=0$ for all
$x\in K$. Whence the argument $a\in N$ satisfies the constraint
$b(x,a)=x^\transp B a=0$ for all $x\in K$, implying $Ba=0$ as claimed.
Moreover, equation (\ref{eq:inter}) can be simplified to
\begin{equation}\label{eq:fin} 
\gamma-b(a,y)+q(a)+d^\transp (y-a)=0.
\end{equation}
Since this holds for all $y\in N$, we get $-b(a,y)+d^\transp y=0$ for
all $y\in N$; hence $-b(a,y)+d^\transp y=(d-aM_y)^\transp y=0$ for all
$y\in N$.  This shows that $Z^d=Z^{aM_y}$.  Finally, substituting
$y=a$ in (\ref{eq:fin}) yields $\gamma=q(a)$, which proves the claim.

Two different elements $\w^{q(a)}X^a Z^{aM_y},\w^{q(a')}X^{a'}
Z^{a'M_y}\in S_Q$ commute, since the symplectic inner product
(\ref{eq:symplectic}) of $(a,aM_y)$ and $(a',a'M_y)$ vanishes as the
matrix $M_y$ is symmetric. As the rank of the matrix $B$ is assumed to
be $k$, there are $n-k$ different vectors $a$ that satisfy the
constraint. Hence the group generated by $S_Q$ has at least $p^{n-k}$
elements.

A projection onto the joint eigenspace $E_1\subseteq \Hilb^{\otimes
n}$ with eigenvalue $1$ of the operators in $S_Q$ generating the group
$S$ is given by
$$
P=\frac{1}{|S|}\sum_{M\in S} M. 
$$
The dimension of $E_1$ is bounded from above by $\dim E_1={\rm tr} P =
p^n/|S|\le p^n/|S_Q|=p^k$. Since $E_1$ contains $Q$ and $\dim Q=p^k$,
this shows that $E_1=Q$. We can conclude that $Q$ is a stabilizer
code.

Omitting the phase factors $\omega^{q(a)}$, we obtain an equivalent
code $Q'$ whose stabilizer group is
\begin{equation}\label{eq:stab_grp2}
S_{Q'}:=\{ X^{\bm{a}}Z^{\bm{a M_y}}
\colon \bm{a}\in \langle B\rangle^\bot=\langle D\rangle\}.
\end{equation}
Here $\langle B\rangle^\bot=\langle D\rangle$ denotes the linear space
of elements $a\in N$ such that $Ba=0$, and the $(n-k)\times n$ matrix
$D$ is formed by a basis of that space. The stabilizer group $S_{Q'}$
gives rise to a symplectic code over $\F_p\times \F_p$ \cite{Rains99}.
This code is generated by the matrix $D(I+\alpha\cdot M_y)$ where
$\{1,\alpha\}$ is a basis of $\F_{p^2}$ over $\F_p$. Additionally, we
have used the isomorphism of $\F_p\times \F_p$ and $\F_{p^2}$ as
vector spaces.

\section{The Graph of a Stabilizer Code}
In \cite{ashikhmin00} it is shown that any stabilizer code that is
defined via a code over $\F_{p^m}$ can be regarded as a stabilizer
code over $\F_p$. Hence it is sufficient to consider stabilizer codes
over a space $\Hilb$ of prime dimension. Those codes correspond to
symplectic codes over $\F_{p^2}$ \cite{Rains99}.

Let ${\cal C}$ be a symplectic code over $\F_{p^2}$ which is generated
by the matrix $X+\alpha Z=:(X|Z)$ where $\{1,\alpha\}$ is a basis of
$\F_{p^2}$ over $\F_p$ and $X,Z\in\F_p^{(n-k)\times n}$. Furthermore,
let ${\cal C}^\bot$ denote the orthogonal code of ${\cal C}$ with
respect to the symplectic inner product on $\F_{p^2}^n$. As ${\cal
C}\le{\cal C}^\bot$, there exists a self-dual code ${\cal D}$ with
${\cal C}\le{\cal D}={\cal D}^\bot\le{\cal C}^\bot$. Identifying
$\F_{p^2}^n$ and $\F_p^{2n}$ and rearranging the coordinates, we can
choose a generator matrix for ${\cal D}$ of the form
\begin{equation}\label{eq:sym_gen_mat}
G':=(X'|Z')=\left(\begin{array}{c|c}
                X&Z\\\hline\widetilde{X}\rule{0pt}{12pt}&\widetilde{Z}
              \end{array}\right).
\end{equation}
The group of isometries of the symplectic space $\F_{p^2}^n$ that
additionally preserve the Hamming weight is given by the wreath
product of the symplectic group ${\rm Sp}_2(p)$ and the symmetric
group $S_n$ \cite{Rains99}. The symmetric group acts on the generator
matrix (\ref{eq:sym_gen_mat}) by simultaneously permuting columns of
$X'$ and $Z'$. The elements of ${\rm Sp}_2(p)$ operate from the right
on the $i^{\text{th}}$ column of the submatrix $X$ and the
$i^{\text{th}}$ column of the submatrix $Z$. On the rows of $G'$
operates the full linear group. Similar to Gau\ss' algorithm, $G'$ can
assumed to be of the form $G'=(I|P)$ \cite{GrasslCRC}. As ${\cal D}$
is self-dual with respect to the symplectic inner product
(\ref{eq:symplectic}), we have $I\cdot P^\transp-P\cdot I=0$, i.\,e.,
$P$ is symmetric.

In the next step we perform column operations in order to obtain a
matrix of the form $(I|C)$ where $C$ is symmetric and all entries on
the diagonal of $C$ are zero. The code generated by $(I|C)$ is
equivalent to ${\cal D}$, and in particular, it contains a subcode
that is equivalent to ${\cal C}$. Let this subcode be generated by
$D\cdot(I|C)$. Furthermore, let $B$ be a $k\times n$ parity check
matrix for the linear code $[n,n-k]$ over $\F_p$ generated by
$D$. Then the matrix
$$
\Gamma=\left(
\begin{array}{c|c}
0&B\\
\hline
B^\transp & C
\end{array}
\right)
$$
is of the form (\ref{eq:Gamma_block}). The entries of $\Gamma$ are
elements of $\F_p$ which can be interpreted as integers modulo $p$. As
$\Gamma$ is symmetric and the diagonal entries are zero, $\Gamma$ is
the adjacency matrix of a weighted undirected graph. Repeating the
arguments of Section \ref{sec:graphcodes} and Section
\ref{sec:stabilizer}, it can be shown that the graphical quantum code
defined by $\Gamma$ is equivalent to ${\cal C}$. Hence, for any
stabilizer code over $\F_{p^m}$ there exists an equivalent graphical
quantum code.

\section{Examples}

\subsection{The stabilizer of a graphical quantum code}
\begin{figure}[tb]
\centerline{\small
\unitlength1pt
\begin{picture}(110,110)
\put(0,0){\epsffile{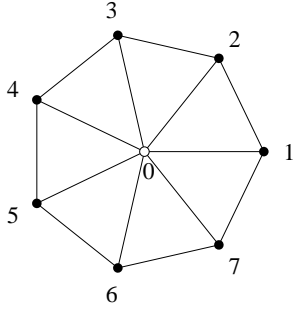}}
\put(56.75,48){\makebox(0,0){0}}
\put(110,55){\makebox(0,0){1}}
\put(89.29, 98.00){\makebox(0,0){2}}
\put(42.76,108.62){\makebox(0,0){3}}
\put( 5.45, 78.86){\makebox(0,0){4}}
\put( 5.45, 31.14){\makebox(0,0){5}}
\put(42.76,  1.38){\makebox(0,0){6}}
\put(89.29, 12.00){\makebox(0,0){7}}
\end{picture}}
\caption{
The wheel $W_7$ with 7 vertices of degree 3 yields a 
$[\![7,1,3]\!]$ QECC.\label{fig:graph7}}
\end{figure}

Consider the highly symmetric graph depicted in Fig.~\ref{fig:graph7}.
This graph is called the wheel $W_7$.  All edges in this graph have the same
weight, hence its adjacency matrix is given by
$$
\Gamma_{W_7}= 
\left(\mbox{\small$
\begin{array}{c|ccccccc}
0&1&1&1&1&1&1&1\\ \hline
1&0&1&0&0&0&0&1\\
1&1&0&1&0&0&0&0\\
1&0&1&0&1&0&0&0\\
1&0&0&1&0&1&0&0\\
1&0&0&0&1&0&1&0\\
1&0&0&0&0&1&0&1\\
1&1&0&0&0&0&1&0
\end{array}
$}
\right),
$$
where we have indicated the block structure as in
eq.~(\ref{eq:Gamma_block}). The $7\times 1$ submatrix $B$ corresponds
to the repetition code, so the matrix $D$
(cf. eq.~(\ref{eq:stab_grp2})) generates the even weight code of
length~$7$. Using the notation of (\ref{eq:sym_gen_mat}), we obtain
$$
\begin{array}{@{}r@{}c@{}l@{}}
G&{}={}&D\cdot(I|M_y)\\[1ex]
&{}={}&
\left(\mbox{\small$\arraycolsep0.9\arraycolsep
\begin{array}{*7c|*7c}
1&0&0&0&0&0&1 & 1&1&0&0&0&1&1\\
0&1&0&0&0&0&1 & 0&0&1&0&0&1&0\\
0&0&1&0&0&0&1 & 1&1&0&1&0&1&0\\
0&0&0&1&0&0&1 & 1&0&1&0&1&1&0\\
0&0&0&0&1&0&1 & 1&0&0&1&0&0&0\\
0&0&0&0&0&1&1 & 1&0&0&0&1&1&1
\end{array}
$}
\right),
\end{array}
$$
where $M_y$ is the lower right $7\times 7$ submatrix of
$\Gamma_{W_7}$. The corresponding additive code over ${\rm GF}(4)$
\cite{CRSS98} is generated by
$$\def\a{\alpha}
G_4:=\left(\mbox{\small$
\begin{array}{*7c}
\a^2&	\a&    0&    0&	   0&	\a& \a^2\\
   0&	 1&   \a&    0&	   0&	\a&    1\\
  \a&	\a&    1&   \a&	   0&	\a&    1\\
  \a&	 0&   \a&    1&	  \a&	\a&    1\\
  \a&	 0&    0&   \a&	   1&	 0&    1\\
  \a&	 0&    0&    0&	  \a& \a^2& \a^2\\
\end{array}$}\right),
$$
where $\alpha$ denotes a primitive element of ${\rm GF}(4)$. The
additive code $C_4=(7,2^6)$ generated by $G_4$ is not ${\rm
GF}(4)$-linear as $G_4$ has rank $6$ over ${\rm GF}(4)$. The weight
distribution of $C_4$ and its dual $C_4^\bot$ are
$$
\mbox{\small$\thickmuskip0.95\thickmuskip
\begin{array}{@{}r@{}c@{}l@{}}
W_{C_4}(x,y)&{}={}&x^7+21 x^3y^4+42x y^6\\[1ex]
W_{C_4^\bot}(x,y)&{}={}&x^7+21 x^4y^3+21 x^3y^4+126x^2y^5+42xy^6+45y^7.
\end{array}
$}
$$
Thus the corresponding stabilizer code has minimum distance~$3$.

\subsection{The graph of a stabilizer code}
\begin{figure}[tb]
\centerline{
\begin{tabular}{cc}
\def\epsfsize#1#2{0.8#1}\epsffile{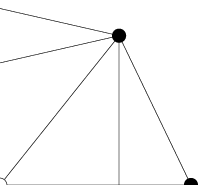}&
\def\epsfsize#1#2{0.8#1}\epsffile{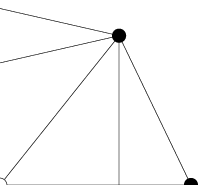}\\
\def\epsfsize#1#2{0.8#1}\epsffile{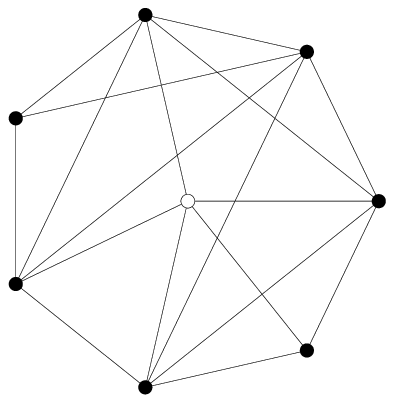}&
\def\epsfsize#1#2{0.8#1}\epsffile{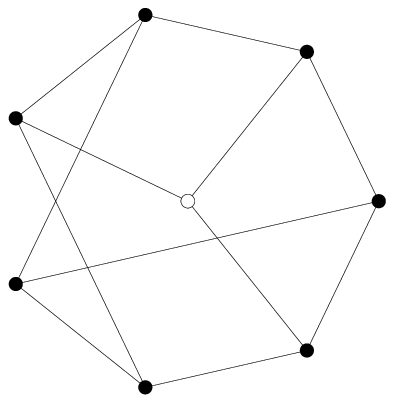}
\end{tabular}}
\caption{Four non-isomorphic graphs which yield graphical quantum
codes that are equivalent to the CSS code $[\![7,1,3]\!]$.\label{fig:hamming_graphs}}
\end{figure}

Consider the CSS code (cf.~\cite{CaSh96,Steane96}) $[\![7,1,3]\!]$
derived from the $[7,4,3]$ Hamming code over $\F_2$. The corresponding
additive code ${\cal C}_7$ is generated by
$$
G=
\left(\mbox{\small$\arraycolsep0.9\arraycolsep
\begin{array}{*7c|*7c}
1&0&0&1&0&1&1\\
0&1&0&1&1&1&0\\
0&0&1&0&1&1&1\\
&&&&&&&      1&0&0&1&0&1&1\\
&&&&&&&      0&1&0&1&1&1&0\\
&&&&&&&      0&0&1&0&1&1&1
\end{array}$}
\right),
$$
which has block diagonal form. In order to obtain a generator matrix
$G'$ for the self-orthogonal code ${\cal D}$ with ${\cal C}\le{\cal
D}\le{\cal C}^\bot$, we have to add a non-zero vector of the
complement of ${\cal C}$ in ${\cal C}^\bot$ to $G$. In our example, we
choose the all-ones vector. Next we transform $G'$ into the form
$(I|C)$ where $C$ is symmetric. We obtain
\begin{equation}\label{eq:normalform1}
\begin{array}{@{}r@{}c@{}l@{}}
(I|C)&{}={}&T\cdot G'\cdot S\\
&{}={}&
\left(\mbox{\small$\arraycolsep0.9\arraycolsep
\begin{array}{*7c|*7c}
1&0&0&0&0&0&0&0&0&1&1&1&0&0\\
0&1&0&0&0&0&0&0&0&1&0&1&0&0\\
0&0&1&0&0&0&0&1&1&0&0&0&1&0\\
0&0&0&1&0&0&0&1&0&0&0&0&1&1\\
0&0&0&0&1&0&0&1&1&0&0&0&0&1\\
0&0&0&0&0&1&0&0&0&1&1&0&0&0\\
0&0&0&0&0&0&1&0&0&0&1&1&0&0
\end{array}$}\right),
\end{array}
\end{equation}
where $S\in {\rm Sp_2(2)\wr S_7}$, and $T\in{\rm  GL}_7(2)$ is
given by
$$
T=\left(\mbox{\small$
\begin{array}{*7c}
1&0&1&1&0&1&0\\
1&0&0&1&0&0&1\\
0&0&0&1&1&1&0\\
0&0&0&1&0&0&0\\
0&0&0&1&1&0&0\\
1&1&0&1&1&0&1\\
1&1&1&1&1&1&1
\end{array}$}\right).
$$
In our example, the matrix $B^\transp$ is given by the last column of
$T$. This leads to the adjacency matrix
$$
\Gamma_{\text{Hamming}}=
\left(\mbox{\small$
\begin{array}{c|*7c}
0&0&1&0&0&0&1&1\\\hline
0&0&0&1&1&1&0&0\\
1&0&0&1&0&1&0&0\\
0&1&1&0&0&0&1&0\\
0&1&0&0&0&0&1&1\\
0&1&1&0&0&0&0&1\\
1&0&0&1&1&0&0&0\\
1&0&0&0&1&1&0&0
\end{array}$}
\right).
$$
Note that the ``normal'' form (\ref{eq:normalform1}) is not unique,
wherefore the corresponding graph is not unique either. In
Fig.~\ref{fig:hamming_graphs} we have depicted four obviously
non-isomorphic graphs which all lead to graphical quantum codes that
are equivalent to the CSS code $[\![7,1,3]\!]$. The lower right graph
is a permuted version of $\Gamma_{\text{Hamming}}$. As in
Fig.~\ref{fig:graph7}, the first node is drawn as an open circle.
Furthermore, none of the graphs reflects the cyclic symmetry of the
quantum code.

\section{Conclusions}
We have shown that any stabilizer code over a finite field has an
equivalent representation as a graphical quantum code.  Unfortunately,
this representation is not unique, neither does it reflect all the
properties of the quantum code. However, the construction of good
quantum codes with the help of graphs is a promising avenue for
further research. It should be noted that independent of this work,
Dirk Schlingemann has also established the equivalence of graphical
quantum codes and stabilizer codes \cite{Schlingemann:2002}.

\section*{Acknowledgments}
The authors acknowledge discussions with D.~Schlingemann and
R.~F.~Werner on early versions of \cite{schlingemann00}. Part of this
work was supported by the European Community under contract
IST-1999-10596 (Q-ACTA) and the Deutsche Forschungsgemeinschaft,
Schwerpunktprogramm QIV (SPP~1078), Projekt AQUA (Be~887/13-2).

\nocite{GKR02}

\end{document}